\documentstyle[12pt,aaspp4]{article}
 
\input{epsf}

\journalid{}{}
\articleid{}{}
\begin{document}
\def\lax    {\ifmmode{_<\atop^{\sim}}\else{${_<\atop^{\sim}}$}\fi}
\def\gax    {\ifmmode{_>\atop^{\sim}}\else{${_>\atop^{\sim}}$}\fi}
\def\gtorder{\mathrel{\raise.3ex\hbox{$>$}\mkern-14mu
             \lower0.6ex\hbox{$\sim$}}}
\def\ltorder{\mathrel{\raise.3ex\hbox{$<$}\mkern-14mu
             \lower0.6ex\hbox{$\sim$}}}
 
\long\def\***#1{{\sc #1}}
 
\title{On the Integrated Spectrum of the X-ray Binaries
and the Origin of Soft X-ray Emission from the Bulge of M31}

\author{Konstantin N. Borozdin and William C. Priedhorsky}

\affil{NIS Division, Los Alamos National Laboratory, Los Alamos, NM 87545;
kbor@lanl.gov, wpriedhorsky@lanl.gov}

\begin{abstract}

Using {\it ROSAT} PSPC data, we have performed several tests aimed 
at understanding the origin of the soft X-ray spectral component 
detected from the bulge of M31.  We find that a significant 
soft component in the spectrum of the bulge 
is spatially correlated with the unresolved X-ray emission
near the core of M31, which is probably a hot interstellar medium
or perhaps a population of multiple faint sources.
For the first time, we extracted the spectrum of this unresolved
emission, by removing point sources dominating the integral 
spectrum of the bulge, and found it to be responsible for the most of
soft excess.  A soft spectral component is not at all needed 
to fit the point source spectrum that remains after subtracting 
the unresolved emission. 
The integral spectra of bright point sources, both inside and
outside of the M31 bulge, can be fitted with a single power-law 
in the {\it ROSAT} band. 
Our analysis rules out the previous suggestion that all bulge 
emission in M31 may be generated by low mass X-ray binaries 
(Irwin \& Bregman, 1999).

\end{abstract} 


\section{INTRODUCTION}

M31, the Andromeda Galaxy, is the closest spiral galaxy and belongs 
to the same morphological type as the Milky Way. 
It provides us with a sample of X-ray sources at a uniform 
and relatively nearby range of $\sim$800 kpc (\cite{stag98}). 
Galactic absorption in the direction of M31 is rather low (\cite{stark92}), 
allowing the study of soft X-ray emission down to about 0.2 keV,
which is not possible for the sources in the bulge or Galactic plane of
our own Galaxy.

Observations with the {\it Einstein} and {\it ROSAT} satellites 
revealed multiple point X-ray sources within M31, 
and also significant unresolved X-ray emission in the bulge 
(\cite{tf91,pfj93,sup97}, hereafter Su97). The origin of the 
unresolved emission remains unknown. 
The possibilities include interstellar hot gas, or multiple faint 
point sources (stars or X-ray binaries) below the detection limit. 
Recently Irwin \& Bregman (1999), hereafter IB99, suggested that 
whole X-ray emission of the bulge can be attributed to low-mass X-ray 
binaries (LMXBs), which are commonly assumed to be responsible for 
the hard (above several keV) X-ray emission of the X-ray faint early-type
galaxies.  Whether the same population of LMXBs might also be the origin
for a significant very soft excess detected with {\it Einstein} 
and {\it ROSAT} in several X-ray faint early-type galaxies 
(e.g. \cite{kim92,pel94,isb00}) remained, however, open and interesting
question to explore.

In this letter we present results of our analysis of {\it ROSAT} PSPC data.
We generated spatially resolved spectra of the bulge of M31 and 
for the first time extracted the spectrum of the unresolved X-ray emission. 
The results presented below demonstrate that the integrated 
spectrum of bright X-ray binaries is significantly harder than 
the spectrum of the unresolved emission in the bulge of M31.

\section{OBSERVATIONS AND DATA REDUCTION}

We analyzed the {\it ROSAT} PSPC observations discussed 
in detail by Su97.  Descriptions of the instrument 
can be found in \cite{asch81,pfef82}, and \cite{asch88}.
We used all observations with a total exposure longer than 10,000 s
that were taken before Oct 14, 1991\footnote{different response matrix 
must be used with data collected after this date, hence it is not 
possible to sum spectra together with previous observations}. 
Most of our results were obtained from observation RP600068N00,
when the central part of M31 was observed(Fig.\ \ref{fig_068}). 
The angular resolution of {\it ROSAT}/PSPC
does not allow a clean separation of point sources from the diffuse
emission in the bulge of M31, because both are concentrated
towards the center of the galaxy. However, we
segregated these two components spatially as best as we could, 
collecting small regions around bright binaries into one spectrum, 
and photons from the areas without bright point-like sources 
into another.  To identify point sources we used the {\it ROSAT} PSPC 
catalog published by Su97.  To optimize the radius of photon 
collection for the point sources in the bulge, where diffuse emission 
is significant, we tried collection radii ranging from 30 to 60 arcsec 
for each source.  As the radius increases we are collecting a great
fraction of the source flux, but also the amount of diffuse contamination
is increasing. We found that the
choice of this radius does not change our conclusions described below.
For X-ray binaries in the bulge, the typical radius was chosen to be 
45 arcsec. Studying the integral spectrum of bright point sources 
outside of the bulge of M31, we collected all photons within 
75 arcsec from each source.

We used only data within a central part of {\it ROSAT}'s field of view, 
that is inside the central ring of PSPC support grid with 
$\sim$ 20 arcmin radius. 
The background was measured in the same area of the detector, but
outside of M31 bulge, in a region devoid of significant point sources 
and with much less unresolved emission than 
the bulge. The same background spectrum was used for all analyses of 
observation RP600068N00,  both for bright point sources, 
and for unresolved emission, except when diffuse emission 
was subtracted from the point source spectrum as a background
(as discussed below).  We experimented with different background regions 
and found that our results do not depend on this choice. The spectral 
differences we discuss in this paper are much stronger 
than the expected background variations.

Various routines from the $ftools$ package (v.5.0) 
were used to generate spectra and other necessary files, 
which were then used for spectral fitting with $xspec$ (v.11.0) package. 
We ignored energy channels below 11 and above 219, 
and rebinned spectra so that each channel contained at least 25 counts.

\section{SPECTRAL TEST RESULTS}

Our choice of spectral model was based on recent spectral studies of
M31 bulge reported by IB99 and \cite{tri99}, hereafter Tr99. Both studies 
revealed the presence of at least two spectral components in the integral 
spectrum of the bulge. The hard component was fitted with bremsstrahlung 
or power law model, while soft component - by a Raymond-Smith or MEKAL
plasma emission model.  We chose to use a power law for the hard component, 
mostly because the reported best-fit temperature of bremsstrahlung 
was well above the {\it ROSAT} upper energy limit. 
Unfortunately, {\it ROSAT} data cannot provide us with tight constraints
for parameters of a two-component model.  Table \ref{tab_c5} presents
several different combinations of parameters which all give reasonably
good fits for the integrated spectrum of bulge emission, including 
all point sources and unresolved emission within
inner 5 arcmin radius around the optical nucleus of M31 
(see e.g. \cite{cdc92} for J2000 coordinates).
The parameters of the fit are strongly correlated.  
There is the expected correlation between power law slope and 
absorption value, and also a correlation of the absorption
with metallicity.  The studies by IB99 and Tr99, which used {\it ASCA}
and {\it BeppoSAX} data, also did not provide
well-constrained fit parameters.  The temperature parameter of soft
component is 0.36--0.38 keV in IB99, but varies between 0.15 and
0.50 for the several fits presented by Tr99.  Metallicity can be as low
as 0.03 (IB99) or as high as 1 (fixed by Tr99). The power law slope
from 1.2--1.3 (IB99) to 1.8--2.2 (Tr99), and absorption parameter 
$N_H$ from 4.4 $\times$ 10$^{20}$ cm$^{-2}$ (IB99) to 10--12 
$\times$ 10$^{20}$ cm$^{-2}$ (Tr99).  As one may see from 
Table \ref{tab_c5}, fraction of emission in the soft component 
depends strongly on fit parameters. Therefore, we need to freeze 
most of the parameters in order
to track the relative contributions of the two components 
as a function of position.

For the most of our analysis we fixed the power law slope at 1.5, 
the absorption column at the Galactic value $N_H$ = 6.73 $\times$ 
10$^{20}$ cm$^{-2}$ (\cite{stark92}), and the metallicity parameter 
at 0.03.  This choice allowed us to get 
acceptable values of $\chi^2$ value in all cases, and study relative
significance of hard and soft components for various regions in the
bulge of M31.  It does not imply, though, that these values have 
a real physical sense.  As illustrated by Table \ref{m31_fits}, one 
might, for example, choose to fix the metallicity at solar value 
and get different fit parameters.  However, our main conclusion 
regarding the difference between the spectra of bright point sources 
and the unresolved emission is not model-dependent.

Using the model described above, we looked at the spectrum 
of the integral bulge emission (including all point sources) 
at various distances from the M31 center 
(less than 2 arcmin, between 2 and 5 arcmin, and between 5 and 8 arcmin). 
The contribution from the soft component decreased with distance
from the center, in correlation with the decrease of 
the diffuse intensity in the image. The soft component fraction 
falls from 0.62 to 0.50 and then to 0.37, as we move from the inner 
2 arcmin to the outermost annulus (Table \ref{m31_fits}).

For the next stage of our analysis, we separated bright point sources 
from unresolved emission.  Overall approach was to draw small circular regions
around each point source listed in Su97.  Then we integrated these regions
into one spectrum (`XRBs'), and integrated the rest of the area into a
second spectrum (`diffuse'). Admittedly, we cannot separate 
point sources and unresolved emission near the core completely 
by this method, because the density of unresolved
emission is maximal near the core of M31, where the concentration of point 
sources is also maximal, however, relative contribution of both kinds of
sources is different in spectra of those two types.  In Table \ref{m31_fits}
we present two spectra for each type, which differ by the area where 
the emission was collected.  We tried to collect unresolved emission from
the area where its density corresponds to average density of the same emission
around bright sources.  This allowed us to subtract `diffuse' spectra from 
`XRBs' spectra in an attempt to get `clean' spectrum of point sources.

\begin{figure}
\epsfxsize=14cm
\epsffile{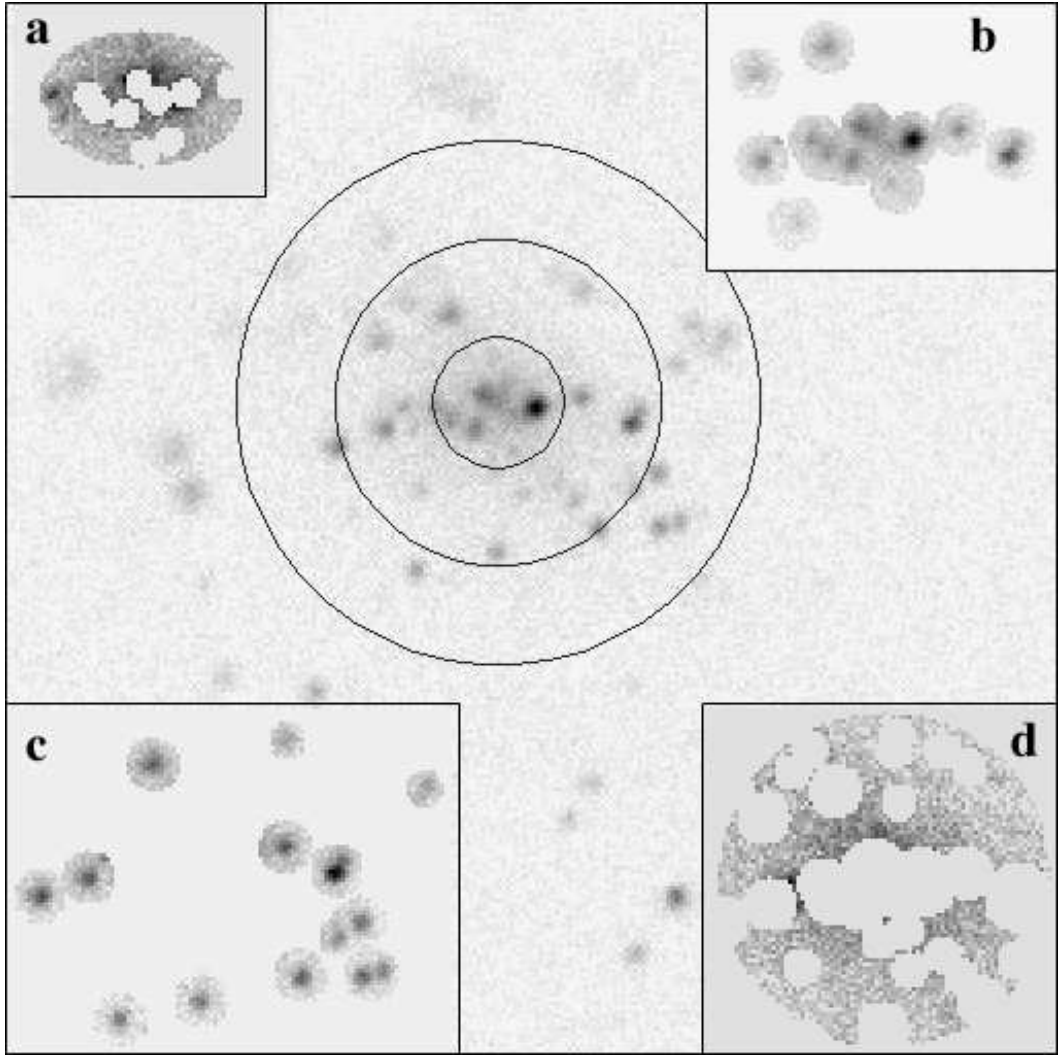} 
\caption{Central part of M31 observed with {\it ROSAT}/PSPC. Concentric
circles are drawn at 2, 5 and 8 arcmin radia from the center.
{\it Inset a:} Unresolved emission within 4x6 arcminute ellipse
after the subtraction of bright point sources (see `diffuse near the core'
spectrum in Table \ref{m31_fits}). Total 7959 counts in the region. 
{\it Inset b:} Bright point sources in the central part of M31
(`XRBs near the core' -- 20969 counts).
{\it Inset c:} Bright point sources in the annulus between 2 and 8
arc from the center (`XRBs in the bulge' -- 11428 counts).
{\it Inset d:} Unresolved emission within 5 arcmin from the center
after the subtraction of bright point sources (`diffuse in the bulge' -- 
6800 counts). Angular scale is the same for the central panel and all insets.
Intensity scale is logarithmic and normalized within each panel to make
faint features more clear.
}
\label{fig_068}
\end{figure}

\begin{figure}
\epsfxsize=17cm
\epsffile{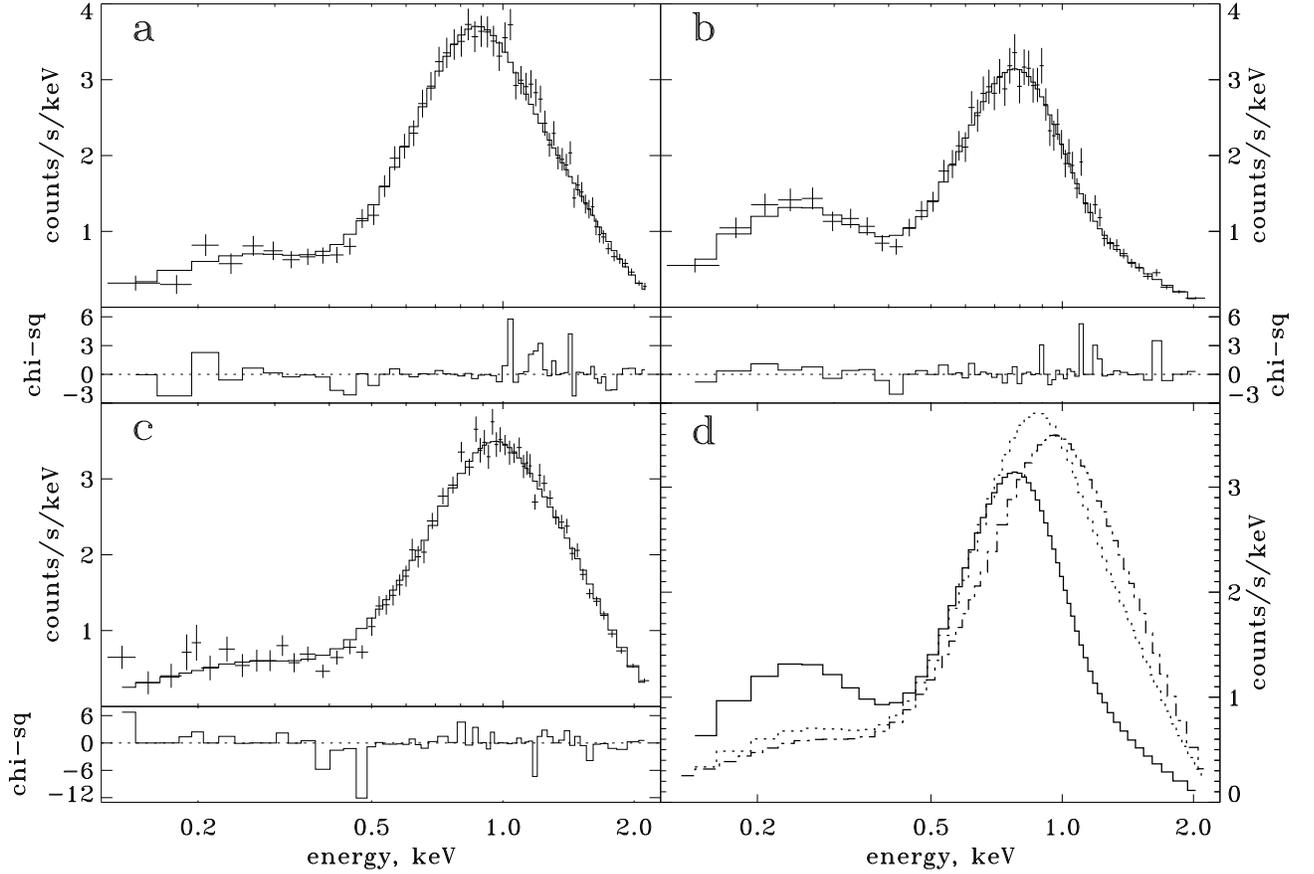} 
\caption{Spectral distributions of {\it ROSAT} counts integrated 
from different regions of M31.
{\it Panel a:} Integral spectrum of XRBs in the bulge (see {\it inset c}
of Fig.\ \ref{fig_068}) fitted with
two-component model described in the text. Model parameters:
$\alpha_{pl}$=1.5, $kT_{e}$=0.32 keV, $Z/Z_{\odot}$=0.2,
$N_H$=7$\times$10$^{20}$ cm$^{-2}$. 
{\it Panel b:} Spectrum of unresolved emission ({\it inset a}
of Fig.\ \ref{fig_068}). Model parameters: 
$\alpha_{pl}$=1.8, $kT_{e}$=0.28 keV, $Z/Z_{\odot}$=1.0,
$N_H$=4.3$\times$10$^{20}$ cm$^{-2}$.
{\it Panel c:} Integral spectrum of 61 bright point sources in M31
outside of central 8' fitted with single power-law component
($\alpha_{pl}$=1.43, $N_H$=7.3$\times$10$^{20}$ cm$^{-2}$).
{\it Panel d:} Comparisons of model presented in {\it panels a-c}.
{\it Solid} line shows unresolved emission, 
{\it dotted} -- XRBs in the bulge,
{\it dash-dotted} -- bright sources of the disk.
Spectra were re-normalized for better visualization.
}
\label{fig_cnts4}
\end{figure}

{\it XRBs and diffuse emission near the core}. For these spectra we 
extracted X-ray photons from the area near the center of M31, where 
the brightest point sources are concentrated and the density of diffuse 
emission is maximal.  The separation of two types of progenitors is
especially challenging in this area.  14 bright point sources from 
Su97 catalog, which are located within 6 x 8 arcmin ellipse around 
the center of M31 (see Inset b of Fig.\ \ref{fig_068}), 
formed XRBs spectrum.  Unresolved emission was collected
over the ellipse 4 x 6 arcmin, but circles of 45 arcsec diameter around each
catalogued source were excluded (Inset a of Fig.\ \ref{fig_068}).  
For `XRBs near the core' the contribution of soft component 
is large (0.48), but significantly lower than for the total 
emission from the same region (0.62).  The soft component dominates 
the spectrum of unresolved emission collected in this area (0.8). 
If `diffuse near the core' spectrum, normalized for area, 
is subtracted from `XRBs near the core', then soft component 
is no longer significant in residual spectrum.
In another analysis we fixed the metallicity parameter at 1.  
While, as expected, the absolute values of soft component fraction 
differ significantly for the different metallicity, 
soft component still contributes significantly more (0.47) 
to the spectrum of the unresolved emission
than to the spectrum of `XRBs' (0.22).  The difference in the spectra
illustrated by Fig. \ref{fig_cnts4} does not depend on our choice of 
model parameters -- it is easily seen in the distribution of counts over 
the spectral band.

{\it XRBs and diffuse emission in the bulge}. Here, we summed the spectra
of point sources and collected an unresolved emission from a larger area, 
where they are more easily separable.  The spectra of 14 bright sources 
that lie within 8 arcmin from the center, but outside of central 2 arcmin, 
were integrated into `XRBs in the bulge'(Inset c of Fig.\ \ref{fig_068}).  
Unresolved emission was collected from inner 5 arcmin, less 45 arcsec 
radii around all point sources listed in Su97 (see Inset d of 
Fig.\ \ref{fig_068}). Soft component is much stronger in the spectrum 
of unresolved emission (80\% of total flux), than in the spectrum 
of `XRBs' ($\sim$20\% of total flux, not detected when spectrum 
of diffuse emission is subtracted).

We see that while all spectra can be satisfactorily fitted 
by the same model, the relative contribution of hard and soft component 
are significantly different for point sources and unresolved emission. 
Because we see the difference even with PSPC data which are seriously confused
in this region, we expect that the difference would be even stronger if
we obtained clean, well spatially resolved spectra.
Furthermore, it motivates us to suggest that hard component 
in the bulge spectrum is generated by X-ray binaries of the bulge, while
the soft component originates from a diffuse source of different nature.
This suggestion can be checked with data coming from Chandra and 
XMM-Newton instruments.

Additionally we analyzed the integrated spectrum 
of other X-ray binaries in M31.  In this case we added together 
spectra of all bright point objects located outside of central 8 arcminutes.
We did not add spectra of the sources marked as foreground stars or
background galaxies in Su97.  Altogether 61 bright 
point sources were added together to the spectrum labeled as
`XRBs in the disk' in Table \ref{m31_fits}. The integrated 
spectrum can be fit by single power-law component with a slope around 1.4 
and absorption slightly higher than the Galactic value 
(Fig.\ \ref{fig_cnts4}). No evidence for soft component was found. 
This spectrum is very close to the spectrum of
`XRBs in the bulge' with diffuse emission subtracted. We need 
to note that all spectra labeled as `XRBs' 
in Table \ref{m31_fits} include all bright compact
sources in M31 and does not exclude supernova remnants or globular clusters.
We suggest that a single power-law spectrum in {\it ROSAT}
energy band is a more suitable 
template for the population of bright compact 
X-ray emitters, than the spectrum of M31 bulge, 
which includes a significant soft component.

\section{DISCUSSION}

Our spatially resolved spectral analysis revealed a qualitative 
difference between the unresolved emission and the bright point
sources in the bulge of M31.  We have performed several tests and 
that gave consistent results.  We found that soft component is
much more prominent in the spectrum of unresolved emission than
in the integral spectrum of bright point sources (presumably LMXBs).

Previous studies were based on the spectrum of integral X-ray emission 
from the bulge, in comparison with spectra for individual sources.
IB99 argued that LMXBs may be responsible for
both hard and soft X-ray components in the spectrum of the M31 bulge.
They suggested furthermore that unresolved emission in the bulge is
composed of the same LMXBs and has the same energy spectrum.
Our analysis shows conclusively that this is not the case. 
The spectrum of the unresolved emission is significantly different 
and hence it must be dominated by the sources of a different nature.
These sources are responsible for most of soft excess observed
from M31.  We cannot rule out the possibility that LMXBs generate 
some of the soft component, but the contribution of this component 
to their spectra is lower than in the unresolved emission.  
The nature of the unresolved emission remains unclear, however, 
we believe that 
the observation of an extended radio source of a similar size 
at the same position in M31 (\cite{hs82}) supports an
interstellar gas hypothesis.  The reality of diffuse emission in M31 
is, in a sense, not a surprise - in our own Galaxy, 
we have long known of diffuse
emission from the Galactic plane down to 3 keV (Valinia \&
Marshall 1998, Kaneda et al. 1999, and references therein).
Although our measurements of M31 did not extend to these energies,
and the soft component that we detected would be hidden by absorption
in our own Galaxy, the Galactic measurements give evidence for thermal
emission from a perhaps hotter phase of the interstellar medium.

Our study suggests that the resolved bright sources in the bulge of M31 
may not be responsible for any of the soft X-ray excess in the spectrum.
In this case, their integral spectrum strongly resembles 
the bright X-ray sources outside of the M31 bulge.  This looks
surprising because it is commonly accepted that disk and bulge sources
belong to different populations of X-ray binaries. It may appear,
however, that LMXBs cannot be distinguished from high-mass X-ray binaries
solely on the basis of their X-ray spectra in {\it ROSAT} band.
After all, the compact objects in both types of binaries are of the 
same nature.

Our conclusion that the soft spectral component mainly emerges from the
unresolved emission does not depend on a quantitative deconvolution
of the bulge spectrum. This is important, because the exact fraction
in the soft component depends strongly on the spectral model
parameters, and cannot be well evaluated with existing data.
Our poor state of knowledge is reflected by the difference
between Tr99 (~15\% soft fraction) and IB99 (38\% or more). The
fraction of the total bulge emission from point sources is also
unclear. Trinchieri and Fabbiano (1991) found that 75\% of the bulge
emission was resolved into 46 point sources, but Primini et al.
(1003) estimated that 45 point sources accounted for only 58\% of
the emission from the inner 5', with the rest unresolved. With these
uncertainties, it is difficult to determine the fraction of soft
emission that comes from LMXBs based on integrated spectra,
motivating our spatially-resolved analysis. New data from XMM-Newton
and Chandra promises to clarify the picture.

The interpretation of our results goes far beyond the case of M31, 
because this galaxy, due to its proximity, often serves as a testbed for 
other early-type galaxies (see e.g. Irwin \& Sarazin, 1998).  
The X-ray spectra of many X-ray-faint galaxies had been found to be 
the sum of a hard component and a very soft component, 
similar to the spectrum of the M31 bulge.  
Our analysis suggests that this soft component comes mostly 
from a hot interstellar gas, or some unknown population of
faint, soft X-ray sources, but not from LMXBs.

\section{ACKNOWLEDGMENTS}

The {\it ROSAT} public data were extracted from the HEASARC
electronic archive operated by the Goddard Space Flight Center (NASA).
The authors are thankful to an anonymous referee and to J.Irwin for
useful comments. KB is glad to acknowledge helpful discussion of
the draft by M.Revnivtsev and A.Vikhlinin.


\newpage
\begin{table}
\caption{Spectral fit parameters of $Raymond+PL$ model for 
the emission from the bulge of M31 (within 5 arcmin radius).\label{bulge_fits}}
\label{tab_c5}
\begin{tabular}{cccccc}
\hline
$\alpha_{pl}$ & $kT_{e}$, keV & $Z/Z_{\odot}$ & $N_H$,$\times 10^{20}$ cm$^{-2}$ & $\chi^2_{\nu}$(d.o.f.)& R$_{soft}^a$\\
\hline
1.0(fixed) & 0.37 & 0.022 &  6.73(fixed) & 1.24 (192) & 0.69\\
1.0(fixed) & 0.37 & 0.032 &  6.06 & 1.19 (191) & 0.66\\ 
1.5(fixed) & 0.33 & 0.029 &  6.73(fixed) & 1.25 (192) & 0.56\\
1.5(fixed) & 0.30 & 0.69 & 4.38 & 1.21 (191) & 0.33\\ 
2.0(fixed) & 0.32 & 0.067 & 6.73(fixed) & 1.28 (192) & 0.31\\
2.0(fixed) & 0.31 & 0.88 & 5.84 & 1.18 (191) & 0.20\\
\hline
\end{tabular}
\par
$^{a}$ -- fraction of soft component in total unabsorbed model flux
in 0.2-2.0 keV band
\end{table}

\begin{table}
\caption{Best-fit parameters for different regions 
of M31.\label{m31_fits}}
\small
\begin{tabular}{lccccccc}
\hline
region & $\alpha_{pl}$ & $kT_{e}$, keV & $Z/Z_{\odot}$ & $N_H^a$ & $\chi^2_{\nu}$(d.o.f.)& $F_{tot}^b$ & $R_{soft}^c$\\
\hline
inner 2 arcmin & 1.5(fixed) & 0.35$\pm$0.02 & 0.03(fixed) & 6.73(fixed) & 1.21 (172) & 10.6 & 0.61$\pm$0.03\\ 
ellipse 6x8 arcmin & 1.5(fixed) & 0.32$\pm$0.02 & 0.03(fixed) & 6.73(fixed) & 1.25 (184) & 16.3 & 0.62$\pm$0.03\\ 
2-5 arcmin ring & 1.5(fixed) & 0.32$\pm$0.02 & 0.03(fixed) & 6.73(fixed) & 1.08 (180) & 10.5 & 0.50$\pm$0.02\\ 
5-8 arcmin ring & 1.5(fixed) & 0.32(fixed) & 0.03(fixed) & 6.73(fixed) & 1.06 (166) & 4.40 & 0.37$\pm$0.05\\ 
\hline
XRBs near the core & 1.5(fixed) & 0.32(fixed) & 0.03(fixed) & 6.73(fixed) & 1.00 (186) & 14.1 & 0.48$\pm$0.02\\ 
 & 1.8 $\pm$ 0.2 & 0.30$\pm$0.03 & 1.0(fixed) & 5.87$\pm$0.55 & 0.89 (183) & 12.0 & 0.22$\pm$0.04\\ 
XRBs near the core$^d$ & 1.49$\pm$0.07 & - & - & 6.73(fixed) & 0.62 (186) & 5.28 & 0.\\ 
XRBs in the bulge & 1.5(fixed) & 0.32(fixed) & 0.03(fixed) & 6.73(fixed) & 1.17 (176) & 6.65 & 0.19$\pm$0.04\\ 
XRBs in the bulge$^d$ & 1.31$\pm$0.05 & - & - & 6.73(fixed) & 1.14 (176) & 4.36 & 0.0\\ 
XRBs in the disk & 1.43$\pm$0.04 & - & - & 7.3 $\pm$ 0.4 & 1.29 (86) & 2.18 & 0.0\\ 
\hline
diffuse near the core & 1.5(fixed) & 0.31$\pm$0.01 & 0.03(fixed) &  6.73(fixed) & 1.06 (141) & 5.76 & 0.77$\pm$0.03\\ 
 & 1.8$\pm$0.35 & 0.28$\pm$0.02 & 1.0(fixed) &  4.6 $\pm$ 1.0 & 0.93 (139) & 5.76 & 0.47$\pm$0.07\\ 
diffuse in the bulge & 1.5(fixed) & 0.31$\pm$0.02 & 0.03(fixed) & 6.73(fixed) & 1.25 (129) & 3.98 & 0.80$\pm$0.05\\
\hline
\end{tabular}
\par
$^{a}$ -- in units of $\times 10^{20}$ cm$^{-2}$\\
$^{b}$ -- total unabsorbed model flux in 0.2-2 keV band, in units of
$10^{-12}$ ergs s$^{-1}$ cm$^{-2}$\\
$^{c}$ -- fraction of soft component in total unabsorbed model flux
in 0.2-2 keV band\\
$^{d}$ -- in  this case the spectrum of diffuse emission
has been taken as a background
\end{table}

\end{document}